\documentclass[sigconf]{acmart}

\AtBeginDocument{%
  \providecommand\BibTeX{{%
    \normalfont B\kern-0.5em{\scshape i\kern-0.25em b}\kern-0.8em\TeX}}}

\setcopyright{acmcopyright}
\copyrightyear{2022}
\acmYear{2022}
\acmDOI{10.1145/3503161.3548357}

\acmConference[Lisbon '22]{Lisbon '22: ACM MultiMedia}{October 10--14, 2022}{Lisbon, Portugal}
\acmBooktitle{Lisbon '22: ACM MultiMedia, October 10--14, 2022, Lisbon, Portugal}
\acmPrice{15.00}
\acmISBN{978-1-4503-XXXX-X/18/06}

\usepackage{amsmath}
\usepackage{amsthm}
\usepackage{booktabs}
\usepackage{algorithm}
\usepackage{algorithmic}
\usepackage{amsfonts}
\usepackage{multirow}
\usepackage{bm}
\usepackage{xcolor}
\usepackage{graphicx}
\usepackage{colortbl}
\usepackage{url}
\usepackage{hyperref}
\usepackage{wrapfig}
\usepackage{caption}
\usepackage{subcaption}

\def\myname{ReLyMe}

\acmSubmissionID{2758}

\begin{document}

\title{ReLyMe: Improving Lyric-to-Melody Generation \\ by Incorporating Lyric-Melody Relationships}


\author{Chen Zhang}
\authornote{Both authors contributed equally to this research.}
\email{zc99@zju.edu.cn}
\affiliation{%
  \institution{Zhejiang University}
  \city{Hangzhou}
  \country{China}
}

\author{Luchin Chang}
\authornotemark[1]
\email{changluchin@gmail.com}
\affiliation{%
  \institution{Zhejiang University}
  \city{Hangzhou}
  \country{China}
}

\author{Songruoyao Wu}
\authornotemark[1]
\email{22021296@zju.edu.cn}
\affiliation{%
  \institution{Zhejiang University}
  \city{Hangzhou}
  \country{China}
}

\author{Xu Tan}
\email{xuta@microsoft.com}
\affiliation{%
  \institution{Microsoft Research Asia}
  \city{Beijing}
  \country{China}
}

\author{Tao Qin}
\email{taoqin@microsoft.com}
\affiliation{%
  \institution{Microsoft Research Asia}
  \city{Beijing}
  \country{China}
}

\author{Tie-Yan Liu}
\email{tyliu@microsoft.com}
\affiliation{%
  \institution{Microsoft Research Asia}
  \city{Beijing}
  \country{China}
}

\author{Kejun Zhang}
\authornote{Corresponding author.}
\email{zhangkejun@zju.edu.cn}
\affiliation{%
  \institution{Zhejiang University}
  \institution{Alibaba-Zhejiang University Joint Institute of Frontier Technologies}
  \city{Hangzhou}
  \country{China}
}

\begin{abstract}

Lyric-to-melody generation, which generates melody according to given lyrics, is one of the most important automatic music composition tasks. With the rapid development of deep learning, previous works address this task with end-to-end neural network models. However, deep learning models cannot well capture the strict but subtle relationships between lyrics and melodies, which compromises the harmony between lyrics and generated melodies. In this paper, we propose \myname{}, a method that incorporates \underline{Re}lationships between \underline{Ly}rics and \underline{Me}lodies from music theory to ensure the harmony between lyrics and melodies. Specifically, we first introduce several principles that lyrics and melodies should follow in terms of tone, rhythm, and structure relationships. These principles are then integrated into neural network lyric-to-melody models by adding corresponding constraints during the decoding process to improve the harmony between lyrics and melodies. We use a series of objective and subjective metrics to evaluate the generated melodies. Experiments on both English and Chinese song datasets show the effectiveness of \myname{}, demonstrating the superiority of incorporating lyric-melody relationships from the music domain into neural lyric-to-melody generation.
\end{abstract}




\maketitle

\section{Introduction}
Music has extremely versatile applications in daily life, commerce, and culture, yet the music composition process can be quite time-consuming and expensive for humans. Recently, the progress of generative models~\cite{sohl2015deep,vaswani2017attention,dai2019transformer} and the growing demand for music have further motivated the development of automatic music composition. Lyric-to-melody generation, one of the most important and common automatic music composition tasks, has attracted increasing attention from both academia and industry. 
Previous works addressed lyric-to-melody generation mainly with neural network models~\cite{bao2019neural,lee2019icomposer,yu2021conditional,sheng2021songmass,ju2021telemelody}. It is difficult for neural network models to capture the strict but subtle lyric-melody relationships that are crucial for the harmony between lyrics and melodies~\cite{guo2021automatic}. Therefore, previous deep learning lyric-melody generation systems face the problem that the generated melodies are dissonant with the lyrics. In this paper, we exploit the relationships between lyrics and melodies to ensure that the generated melodies are in harmony with the corresponding lyrics. 

In the last few decades, musicians and linguists have discovered that satisfying some musical relationships between lyrics and melodies is essential to guarantee the overall harmony in songwriting~\cite{yu2008study,schellenberg2013realization,nichols2009relationships}. Any song, regardless of the genre, requires its composer to examine the interplay between the melodies and the lyrics. When some lyric-melody relationships have not been met, the songs will become dissonant, even leading to semantic misunderstandings. This kind of dissonance can be alleviated by adjusting the melodies to the lyrics or by adapting the lyrics to be consistent with the melodies on the basis of keeping the two in their own laws. 
Our \myname{} is applied to lyric-to-melody generation systems and ensures the harmony between lyrics and melodies by tuning the generated melodies. However, our summary of the lyric-melodic relationships also applies to the melody-to-lyric generation.

In this paper, we introduce the principles of lyric-melody relationships based on the research from musicians and linguists. 
The relationships are from three aspects: 1) tone, the pitch flow of the generated melodies should be in line with the tone of the lyrics; 2) rhythm, the rhythm of the generated melodies need to match with the rhythm of the lyrics, consisting of the strong/weak positions and the pause positions; and 3) structure, the repetition patterns in generated melodies should echo the repetition in lyrics. Based on these principles, we formulate the rewards that influence the probability of the predicted token at each decoding step to integrate the lyric-melody relationships into neural network models. 

Most previous works on lyric-to-melody generation have evaluated the melody itself alone, with limited consideration (only as a little part of the subjective evaluation) to evaluate the lyrics and melody as a whole. Considering that the melody is generated conditioned on the given lyrics, we introduce additional evaluation metrics to evaluate the relationships between lyrics and generated melodies.
To verify the effectiveness of \myname{}, we apply it in the state-of-the-art end-to-end and multi-stage lyric-to-melody generation systems respectively (SongMASS~\cite{sheng2021songmass} and TeleMelody~\cite{ju2021telemelody}) and conduct the experiments on both English and Chinese song datasets. Both the objective and subjective evaluation results indicate that \myname{} can significantly reduce the lyrical and melodic dissonance of the original lyric-to-melody generation systems, thus further improving the overall quality of the generated melodies combined with the corresponding lyrics. The improvements on both two systems indicate that \myname{} is flexible to be applied in different lyric-to-melody generation systems. Moreover, the method analyses verify the effectiveness of the detailed design of principles and the constrained decoding mechanism in \myname{}. Some demos and our codes can be found in \url{https://ai-muzic.github.io/relyme}.

To sum up, our main contributions are as follows:
\begin{itemize}
  \item We formulate the relationships between lyrics and melodies from three aspects (tone, rhythm, and structure) based on the music theory summarized by musicians.
  \item We propose \myname{}, a method that exploits lyric-melody relationships during the decoding process of neural lyric-to-melody generation systems to ensure harmony between generated melodies and corresponding lyrics.
  \item We conduct experiments on both English and Chinese song datasets and evaluate the generated melodies with objective and subjective evaluation. The results demonstrate the effectiveness of \myname{} in guaranteeing the harmony between lyrics and melodies and show the significance of incorporating lyric-melody relationships into lyric-to-melody generation.
\end{itemize}

\section{Background}
\subsection{Lyric-to-Melody Generation}
Early lyric-to-melody generation works are mainly based on musical rules~\cite{monteith2012automatic} or statistical methods~\cite{fukayama2010automatic,fukayama2012assistance,long2013t}.
\cite{fukayama2010automatic} regarded lyric-to-melody generation task as an optimal-solution search problem under the constraints of lyrics. 
\cite{monteith2012automatic} proposed a rule-based rhythm generation module and then used an n-gram model to predict pitch.
\cite{long2013t} built a melody composer based on the lyric-note relationships learned by frequent pattern mining.
These methods often require much human effort to design the specific rules and suffer from the problem that generated melodies are monotonous.

Recently, with the rapid development of neural networks and generative models~\cite{vaswani2017attention,huang2020pop}, more and more researchers put their eyes on building automatic lyric-to-melody generation systems based on deep learning methods. 
\cite{lee2019icomposer} created melodies to accompany given text with two sequence-to-sequence models to generate the pitch and duration separately without considering other elements in melodies, such as silent intervals.
\cite{bao2019neural} and \cite{yu2021conditional} trained end-to-end models to generate melodies based on lyrics.
These methods usually require lots of paired data to train a robust lyric-to-melody generation model.
\cite{sheng2021songmass} leveraged unpaired data by pre-training to cope with the lack of paired data. 
\cite{ju2021telemelody} proposed a two-stage generation pipeline based on musical templates to improve data efficiency and generation controllability.
These deep learning methods implicitly model the relationships between lyrics and melodies, however, the subtle lyric-melody relationships are hard to fully capture by neural network models. Once the generated melody violates the criterion of the relationship with corresponding lyrics, it results in the dissonance between lyrics and melodies. In this paper, we propose \myname{} to leverage lyric-melody relationships from music theory in lyric-to-melody generation systems to improve the performance.

\subsection{Lyric-Melody Relationships in Songwriting}
Earlier, musicians and linguists have studied the relationships between lyrics and melodies in the human songwriting process that influence the overall harmony~\cite{yu2008study,greene2010automatic,bishop2012perfect,schellenberg2013realization}. They analyzed and summarized these relationships mainly from the perspective of music theory and songwriting laws.
\cite{yu2008study} studied the lyric-melody relationships in Mandarin songs according to rules summarized during the human songwriting process, but no quantitative formulation was given.
\cite{schellenberg2013realization} summarized and formulated the pitch relationships between lyrics and melodies.
In recent years, some researchers start to obtain lyric-melody relationships by statistical methods~\cite{nichols2009relationships}. 
\cite{ma2021ai} built a music structure analyzer to derive the musical structure from melody and generated lyrics under the guidance of musical structures. 
\cite{guo2021automatic} tackled the song translation problem by involving alignments of pitch, rhythm, and length, while it only considered the tonal languages. 
These works only considered the lyric-melody relationships from limited aspects or only apply to certain types of languages. In this paper, we discuss the relationships between lyrics and melodies comprehensively from three aspects and propose \myname{} which can be applied in different languages based on lyric-melody relationships.

\subsection{Constrained Generation}
Most neural generation tasks for discrete tokens such as machine translation, automatic speech recognition, and melody generation, are free generation without constraints. However, there is a need for some scenarios to involve constraints to guide the generation\cite{hokamp2017lexically,lakew2019controlling,saboo2019integration,li2020rigid,zou2021controllable,guo2021automatic}.
Some works added constraints in the training stage,
\cite{lakew2019controlling} investigated two methods for biasing the output length with a transformer architecture: length-ratio class or length embedding; \cite{li2020rigid} designed symbol sets and adds format embedding when training.
Other works, on the other hand, guide the inference stage with constraints, \cite{hu2019improved} used a dictionary to control the beam search; \cite{saboo2019integration} applied re-ranking after beam search with the combined score. In this paper, we choose to add constraints during the decoding and make a comparison with adding constraints after inference or in the training stage.

\section{Method}
\myname{} takes effect upon current neural lyric-to-melody generation systems by incorporating relationships between lyrics and melodies.
Considering that the generation process of most current neural network lyric-to-melody generation systems is auto-regressive, \myname{} involves lyric-melody relationships into neural network models at each decoding step. Specifically, during inference, \myname{} adjusts the probabilities of each decoding step to influence the current generated token, and further to affect the subsequent tokens.

Denote the training lyric-melody pair as $\{(\bm{x}, \bm{y}) \in (\bm{\mathcal{X}}, \bm{\mathcal{Y}})\}$. 
The original score function of beam search in generation system at $i^{th}$ decoding step is:
\begin{equation}
\begin{aligned}
    s_i = logP(y_i|y_{0:i-1}, x).
\end{aligned}
\end{equation}
At each decoding step, the constrained decoding mechanism adds the reward $R_{cons}$ directly to current score.
$R_{cons}(y_{0:i}, x)$ denotes the total reward at $i^{th}$ decoding step under the musical constraints, which consists of three parts: 1) the tone reward $R_t(y_{0:i}, x)$, 2) the rhythm reward $R_r(y_{0:i}, x)$, and 3) the structure reward $R_s(y_{0:i}, x)$.
Thus, the score function in constrained decoding mechanism can be described as:
\begin{equation}
    s'_i = logP(y_i|y_{0:i-1}, x) + R_{cons}(y_{0:i}, x).
    \label{eq:score_cons}
\end{equation}
Then $s'_i$ is used in beam search as the new score function.

\begin{figure}[!t]
    \centering
    \includegraphics[width=0.5\textwidth]{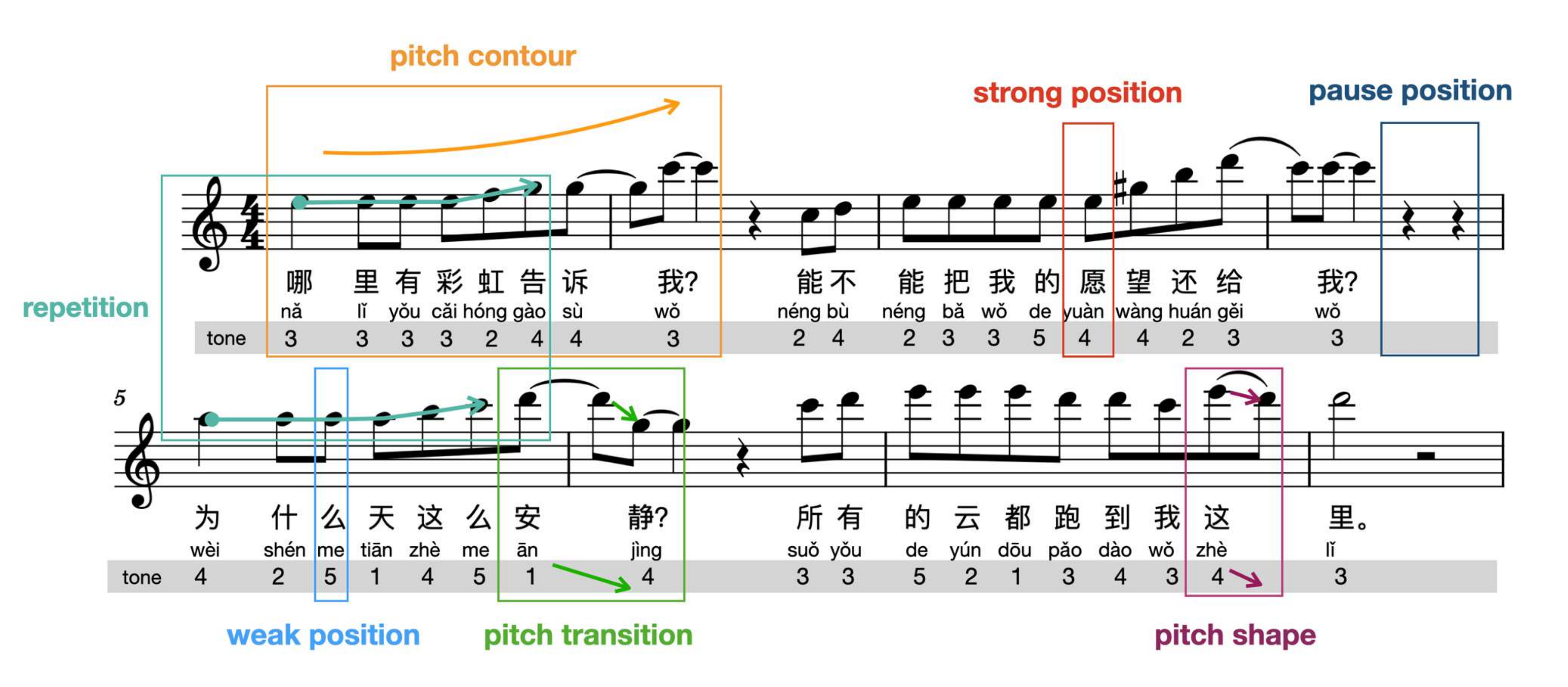}
    \caption{The music sheet of a Mandarin song ``Rainbow'' by Jay Chou. The colorful boxes illustrate the lyric-melody relationships we formulate.}
    \label{fig:case}
\end{figure}

\begin{figure}[!t]
    \centering
    \includegraphics[width=0.4\textwidth]{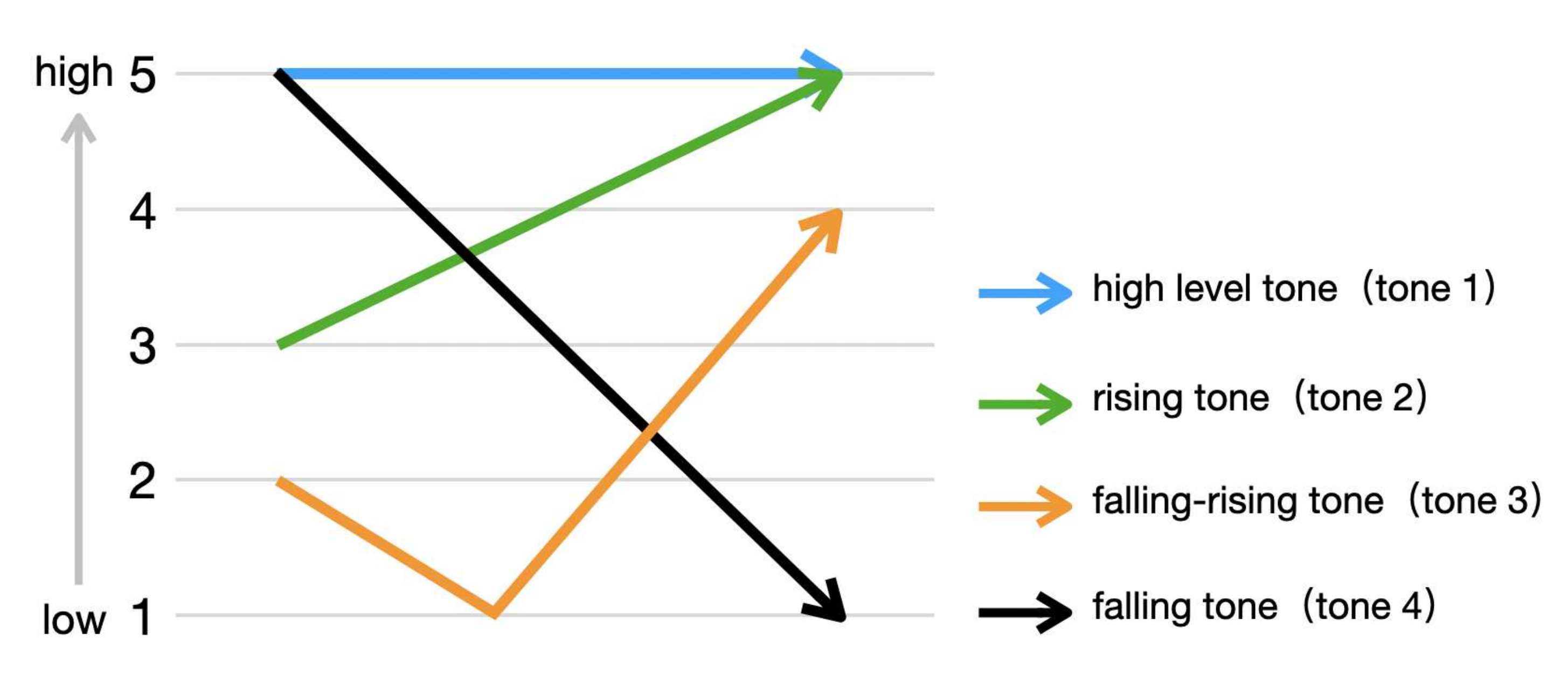}
    \caption{The pitch shape of four main tones in Mandarin.}
    \label{fig:tones}
\end{figure}

We define the above three rewards based on the relationships between lyrics and melodies.
The lyric-melody relationships mean the combining relationships between the rules of lyrics and the rules of the melodies within the scope of unified content, which control the interplay between the lyrics and melodies. According to~\cite{yu2008study} which described the lyric-melody relationships of traditional Chinese folk music from the view of composer or musicians, these combining relationships are reflected in tone, rhythm, and structure. We follow these three aspects and extend them to different types of languages (tonal languages like Mandarin, ``stress accent'' languages like English). An example shows the principles we discuss below can be found in Figure~\ref{fig:case}. 


For a lyric sequence $\bm{x}$, the corresponding syllable sequence is $\bm{s} = \{s_1, ..., s_n\}$, and the tone sequence is $\bm{t} = \{t_1, ..., t_n\}$. The melody $\bm{y}$ consists of a note sequence $\bm{n} = \{n_1, ..., n_m\} (m \geq n)$ and one syllable can be matched with one note or multiple notes. In the following subsections, we introduce the principles and define the rewards (used in Equation~\ref{eq:score_cons}) based on lyric-melody relationships from tone, rhythm, and structure respectively. And then we show the pipelines when applying \myname{} to different lyric-to-melody generation systems.

\begin{figure}[tbp]
	\centering
	\subfloat[A bad case that breaks pitch shape principle. The word ``d\.i f\=ang'' of the lyrics means ``place'', but it sounds like ``d\'{i} f\=ang'', which means ``enemy''.]{\label{fig:case_pitch_shape}\includegraphics[width=0.5\textwidth,trim={0cm, 0cm, 0cm, 1.4cm},clip=true]{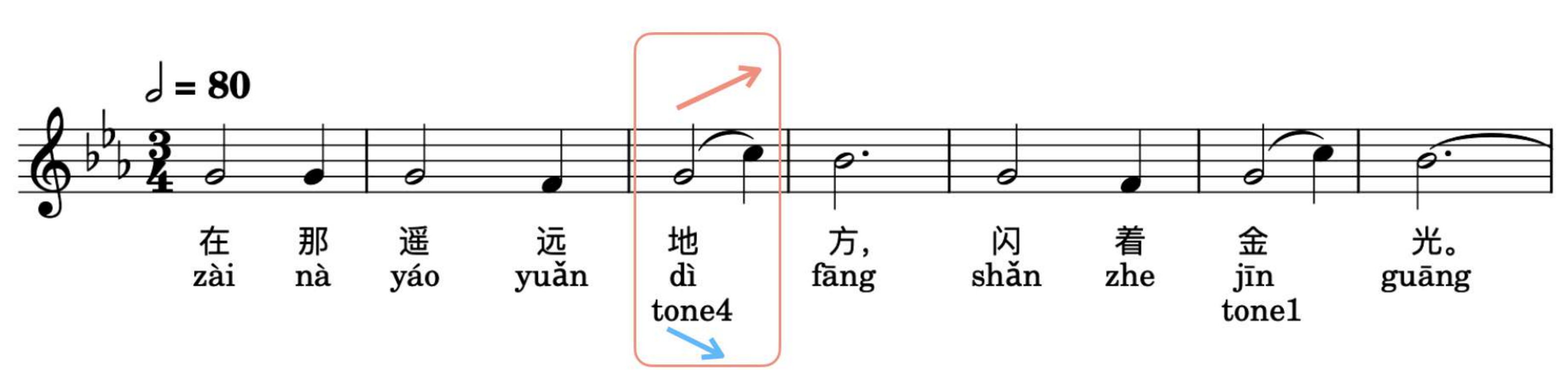}}\\
	\subfloat[A bad case that breaks pitch transition principle. The ``d\=eng t\v{a}'' of the lyrics means ``lighthouse'', but it sounds like ``d\v{e}ng t\=a'', which means ``waiting for him''.]{\label{fig:case_pitch_transition}\includegraphics[width=0.5\textwidth]{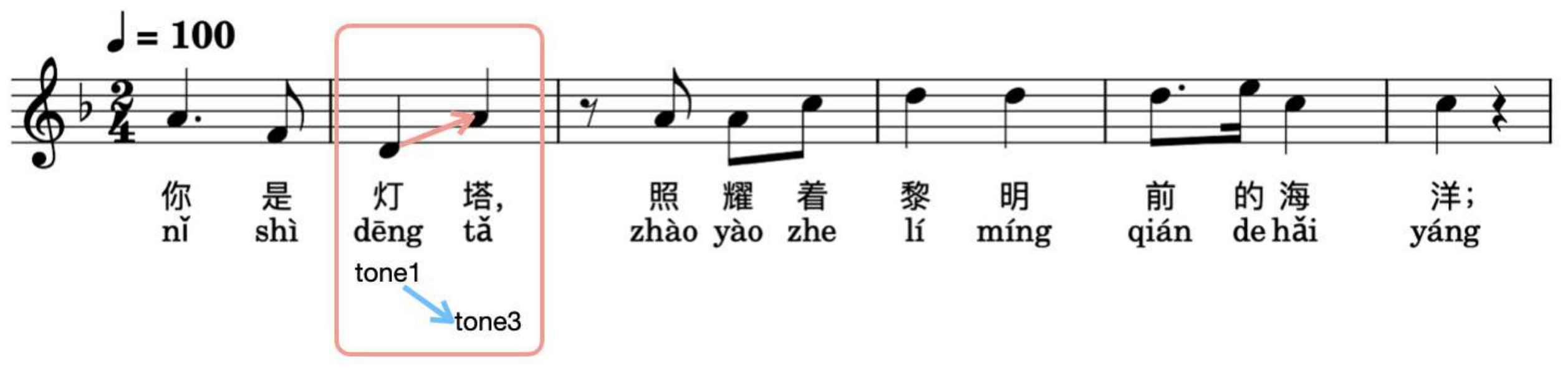}}\\
	\caption{Bad cases of tone principles.}
	\vspace{-0.4cm}
\end{figure}

\begin{figure}[tbp]
    \centering
    \includegraphics[width=0.5\textwidth]{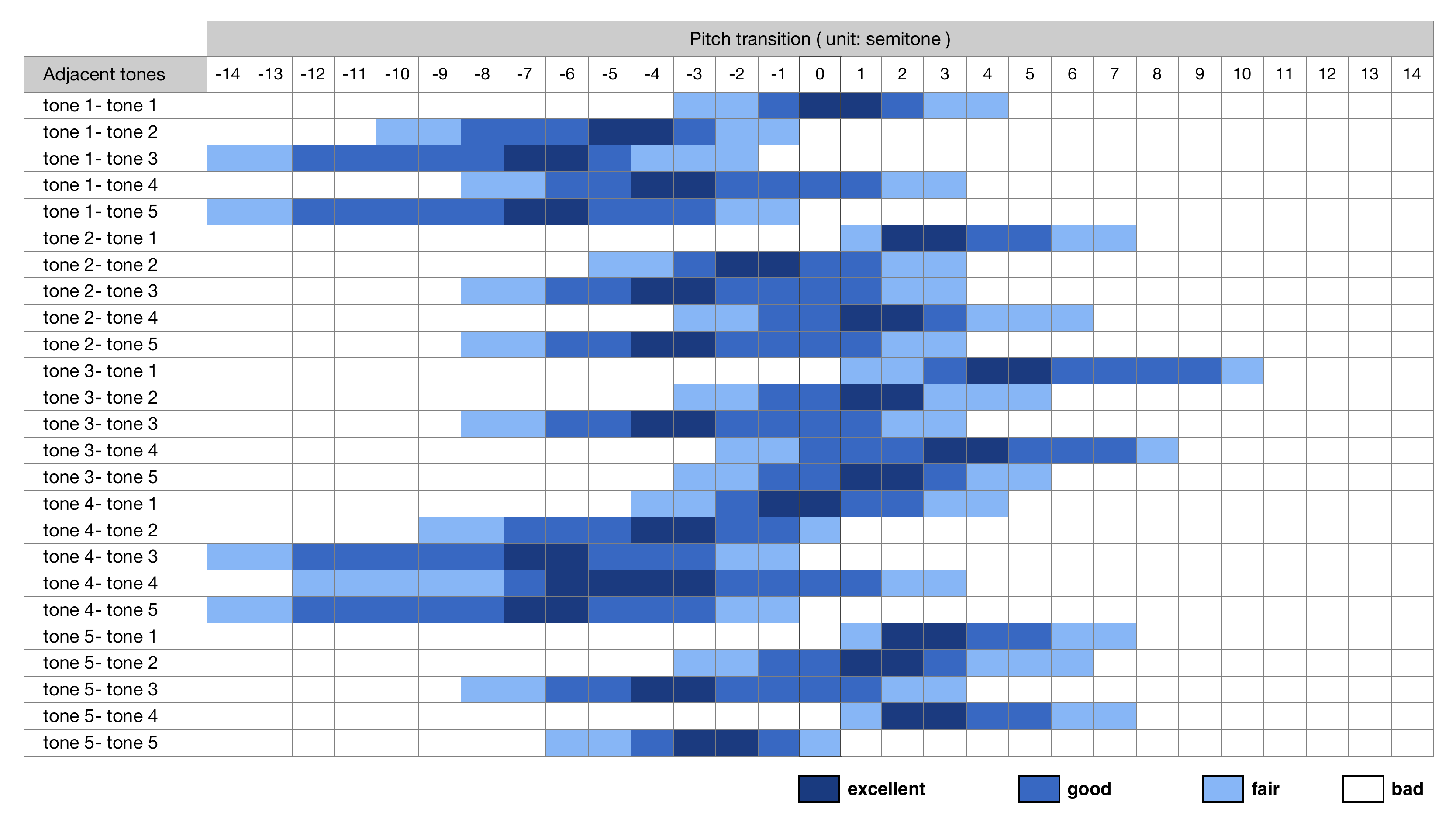}
    \caption{The degree of harmony (excellent, good, fair, bad) between each pitch pair (melody) and each tone pair (lyric) in Mandarin. For each pitch pair, we calculate the pitch transition (pitch difference) of adjacent notes. }
    \label{fig:pitch_transition}
\end{figure}


\subsection{Tone}
\label{sec:tone_formulation}
When the lyrics are sung with given melodies, the pitches of the lyrics are determined by the melodies instead of the intrinsic tone or stress~\cite{zhang2021pdaugment}. Aligning the pitch of the melody with the original pitch of the lyrics is essential for reducing semantic misunderstandings when performing songs~\cite{yu2008study,guo2021automatic}.
About 60\% of languages have tone~\footnote{https://en.wikipedia.org/wiki/Tone\_(linguistics)} (called tonal languages)~\cite{yip2002tone}, such as Chinese, Vietnamese, Thai.
Tone is the pitch information in tonal languages that helps distinguish words with the same spelling but different meanings. In most tonal languages, a single tone is assigned to a syllable.

In order to align the melodies with the lyrics at the tonal level, we consider the following three granularities and define the corresponding rewards respectively.
Denote the pitch sequence corresponding to note sequence as $\bm{p} = \{p_1, ..., p_n\}$. 

\subsubsection{Pitch Shape of a Single Tone}
\ 
\newline
For each syllable in lyrics, when it corresponds to more than one notes in the melody, the predefined shape of the tone should coincide with the pitch flow of the corresponding note group. Take Mandarin as an example, which contains four main tones (high level tone, rising tone, falling-rising tone, falling tone, denoted as tone1, tone2, tone3, and tone4) and a light tone (unstressed syllable pronounced without its original pitch in Chinese pronunciation, denoted tone5)~\cite{hu2017sungwords}. 
The predefined shapes of four main tones are shown in Figure~\ref{fig:tones} and an example that breaks this principle can be found in Figure~\ref{fig:case_pitch_shape}. 

Thus, the pitch shape reward is only considered when more than one notes $\{n_j, ..., n_{j+l}\}$ are assigned to a syllable $s_i$.
Each tone of a syllable in tonal languages has its intrinsic pitch shape, when the pitch shape of the single tone is matched with that of corresponding notes, we assign the reward $R_{ps}(y_{0:i}, x) = 1$, otherwise, the reward is 0.

\subsubsection{Pitch Transition between Adjacent Tones} 
\
\newline
For a pair of adjacent syllables within the same sentence, the pitch flow of the corresponding adjacent notes needs to be in line with the pitch transition between tones, otherwise may lead to semantic misunderstanding like Figure~\ref{fig:case_pitch_transition}. According to the practice in natural speech, we measure the degree of harmony (excellent, good, fair, poor) for the pitch difference between neighboring notes given each adjacent tone pair.

Given each adjacent tone pair $\{t_{i-1}, t_i\}$ in tonal languages, the pitch difference between the corresponding adjacent notes $\{n_{j-1}, n_j\}$ will be classified into four different harmony degrees (excellent, good, fair, bad) according to linguists and musicians (such as Mandarin in Figure~\ref{fig:pitch_transition}~\cite{shao2007general,wang2015tone}). We assign a reward $R_{pt}(y_{0:i}, x)$ to $i^{th}$ decoding step based on the harmony degree of the pitch difference $\Delta p_j = p_j - p_{j-1}$ given the corresponding tone pair $(t_{i-1}, t_i)$. In practice, we assign the reward of 3, 2, 1, 0 when $\Delta p_j$ belongs to the harmony degree of excellent, good, fair, and bad respectively. 

\subsubsection{Pitch Contour of a Whole Sentence} 
\
\newline
Lyrics, as a form of text, have their inherent intonation for each sentence to express different emotions and meanings. The pitch contour of a whole sentence depends on the corresponding melody when performing a song, so the pitch direction of the melody needs to be matched to the inherent intonation of the sentence to ensure that the correct meaning is conveyed. For example, an interrogative tends to align with a rising melody (see the orange box of Figure~\ref{fig:case}).

Assume that $\{t_{i-l}, ..., t_{i}\}$ is the tone sequence of a sentence, we calculate the reward of pitch contour $R_{pc}(y_{0:i}, x)$ according to the inherent intonation of the sentence and the pitch contour of corresponding melody. That is to say, when the inherent intonation and the pitch contour are consistent, the reward is set to be 1, otherwise is 0.

The total tone reward $R_t(y_{0:i}, x)$ is the sum of the above three rewards.
For tonal languages, the pitch of the melody is aligned with the tone of the lyrics, and all of the following three granularities are considered. While for ``stress accent'' languages such as English, which do not have tones, we only consider the pitch contour relationship.

\subsection{Rhythm}

\begin{figure}[tbp]
	\centering
	\subfloat[A bad case that assigns an auxiliary word to downbeat. The question auxiliary ``ma'' is heard as ``m\=a'', which means mother.]{\label{fig:case_strong_weak}\includegraphics[width=0.5\textwidth,trim={0cm, 0cm, 0cm, 1.2cm},clip=true]{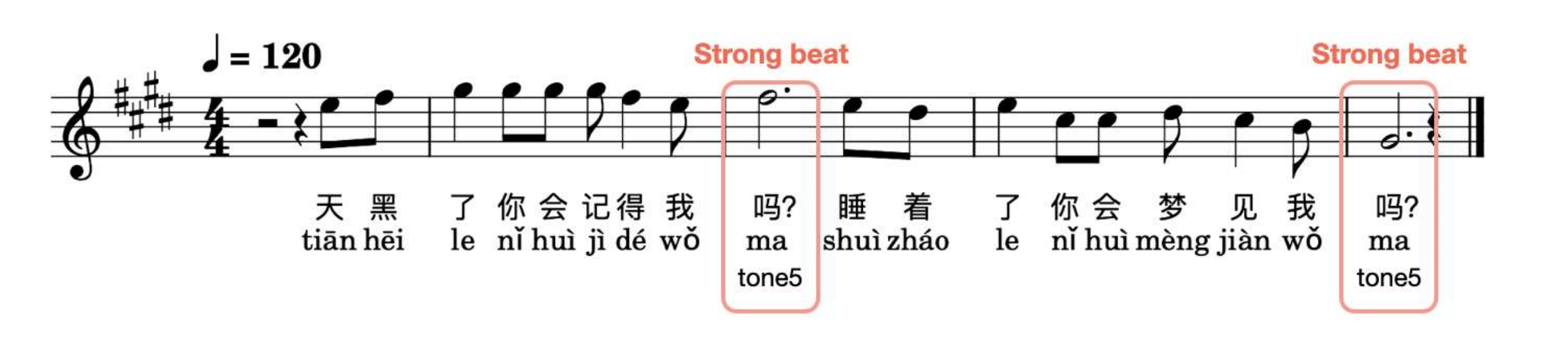}}\\
	\subfloat[A bad case that breaks the principle of pause positions. There is a pause caused by long notes inside a word (red arrow), while an appropriate pause should be at the blue arrow. This mismatch spoils the coherence of the song.]{\label{fig:case_pause}\includegraphics[width=0.5\textwidth]{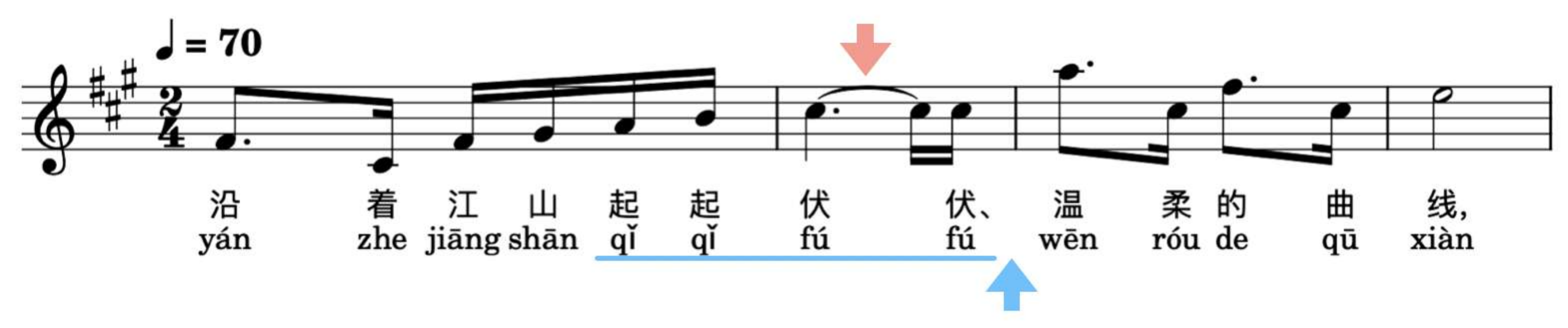}}\\
	\caption{Bad cases of rhythm principles.}
\end{figure}

We consider the rhythmic relationship in two dimensions: 1) intensity at a certain moment, reflected in strong and weak positions; 2) temporal alignment, reflected in pause positions. The total rhythm reward $R_r(y_{0:i}, x)$ is the sum of the following two rewards.

\subsubsection{Strong/Weak Positions}
\
\newline
Both melodies and lyrics have the "strong" and "weak" elements, and the movement marked by the regulated succession of these elements is described as rhythm. A songwriter tends to align downbeats or accents with stressed syllables or important words~\footnote{https://en.wikipedia.org/wiki/Prosody\_(music)}. Aligning the rhythm of melody and lyrics helps convey meaning, express empathy and render emotion. The strong and weak positions in melody lines could be determined by multiple factors~\cite{yu2008study}: pitch, duration, syncopation, meter, and so on. During automatic melody generation, for simplicity, the strong and weak beats can be derived from the predefined time signature~\cite{benward2003introduction}. In lyrics, on the other hand, the strong and weak positions are mainly influenced by oral speaking habits, specific semantics, intent of speakers, and expressed emotions. For example, in spoken language, the auxiliary words tend to be pronounced lightly, while the keywords in the lyrics are often regarded as accents, which are pronounced louder to attract the attention of listeners and emphasize the elements. The strong and weak positions of melodies are aligned with those of lyrics to ensure harmony of performance and correct meaning (a bad case is Figure~\ref{fig:case_strong_weak}). Specifically, the keywords often correspond to the strong positions of the melody while auxiliary words should be placed on the weak positions.
When the strong/weak positions of lyrics and melody match, the reward $R_{sw}(y_{0:i}, x)$ is assigned to 1, otherwise 0.

\subsubsection{Pause Positions}
\
\newline
Pauses in speech divide the entire sentence into several parts with separate meanings (called prosodic phrases) to help the listener understand. Sentences with the same content but different pause positions can have completely different meanings, and the division of prosodic phrases that defies common sense can lead to confusion for listeners. In melody, the pauses are caused by either~\textit{REST} notes~\footnote{https://en.wikipedia.org/wiki/Rest\_(music)} or long notes~\cite{lerdahl1983generative} and split the melody into musical phrases, which make complete musical sense when heard on their own~\cite{frazier2006prosodic}. When the lyrics are sung with given melodies, the pause positions are no longer influenced by the spoken idiom but are controlled by the pauses in the melody. A phenomenon called ``broken phrases'' occurs if syllables of the same prosodic phrase are assigned to different musical phrases. This phenomenon impairs the harmony between the melody and the lyrics and makes the lyrics hard to understand for listeners. Thus, inappropriate \textit{REST} notes or long notes can be detrimental to the music. We summarize the following cases which lead to the dissonance and semantic misunderstanding (examples can be found in Figure~\ref{fig:case_pause}): 1) \textit{REST} notes are located inside the prosodic phrase; 2) long notes are assigned to non-final syllables of a prosodic phrase; 3) there are neither \textit{REST} notes nor long notes at the end of the sentence.

By word segmentation, we can obtain the positions that split words and sentences. The pause positions in melodies need to occur at those ``split'' positions, otherwise prosodic phrases will be broken. Besides, there tends to be a corresponding pause in melody at the position which splits two sentences of lyrics. Thus, the reward $R_{ps}(y_{0:i}, x) = 0$ when the pause in a melody is not at a ``split'' position in corresponding lyrics or there is no pause in a melody between two sentences in the lyrics. Otherwise, $R_{ps}(y_{0:i}, x) = 1$.

\subsection{Structure}

\begin{figure}[!t]
    \centering
    \includegraphics[width=0.5\textwidth]{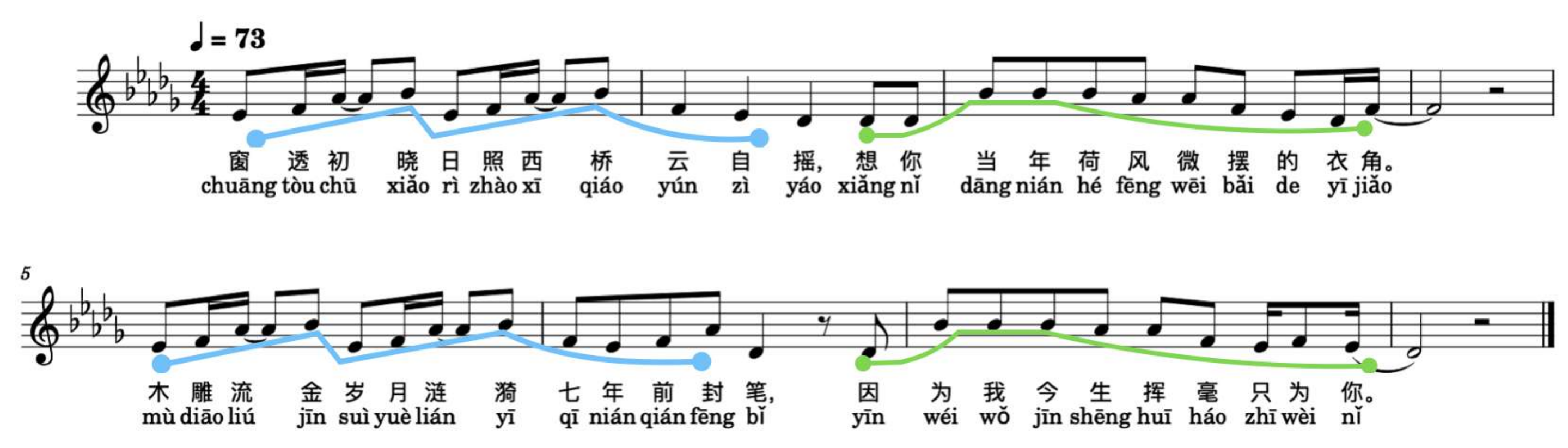}
    \vspace{-0.5cm}
    \caption{An example that shows the structure relationships between lyrics and melodies. The $3^{rd}$ sentence repeats the $1^{st}$ sentence (see blue lines), and the $4^{th}$ sentence repeats the $2^{nd}$ sentence (see green lines).}
    \label{fig:case_structure}
    \vspace{-0.5cm}
\end{figure}

At the full song level, the structure of lyrics and melody are constrained and influenced by each other.
In music, \textit{form}\footnote{https://en.wikipedia.org/wiki/Musical\_form} refers to the structure of a musical composition or performance, including repetition, transition and development~\cite{zhang2021structure}. 
To better convey emotion and show artistry, lyrics and melodies which belong to the same musical form would share the same characteristics like rhyme (in lyrics) or motives (in melodies). Specifically, in lyric-to-melody generation, the melody segments that correspond to the lyrics of the same structure tend to have similar pitch flows (see Figure~\ref{fig:case_structure}).

After the lyrics structure analysis, we can obtain a matrix $M_{struct}$ which indicates the structure of the lyrics. $M_{struct}^{ij} = 1$ if $i^{th}$ note and $j^{th}$ note are at the corresponding positions in the phrases of the same structure and $M_{struct}^{ij} = 0$ otherwise. 
Thus, we assign the structure reward $R_s(y_{0:i}, x) = 2$ if the pitch difference at $i^{th}$ note is equal to that of $j^{th}$ note (i.e. $\Delta p_i = \Delta p_j$ where i and j satisfy $M_{struct}^{ij} = 1$ and $j<i$), that is the pitch sequences of two sentences are equal after overall shifting. And we assign $R_s = 1$ if $\Delta p_i - \Delta p_j$ is a multiple of 12. Otherwise, the structure reward is 0.


\subsection{Applications to Different Lyric-to-Melody Generation Systems}
\myname{} is flexible to be applied to different lyric-to-melody generation systems, and in this work, we apply \myname{} in two systems: 1) TeleMelody~\cite{ju2021telemelody}, a two-stage lyric-to-melody generation system with musical templates; and 2) SongMASS~\cite{sheng2021songmass}, an end-to-end lyric-to-melody generation system with a pre-training process. These two are state-of-the-art systems in end-to-end and multi-stage lyric-to-melody generation systems respectively.

\noindent{\textbf{For TeleMelody.}}
TeleMelody consists of a lyric-to-template module and a template-to-melody module, using the template as the bridge to narrow the gap between lyric and melody, as well as improving the generation controllability by adjusting the template during the generating process. \myname{} adjusts both the lyric-to-template module and template-to-melody module, and guides them to generate templates or melodies which obey the principles.
During the lyric-to-template module, we adjust the score of each step in beam search with the rhythmic constraints:
\begin{equation}
\begin{aligned}
    logP_{l2t}(\bm{z}|\bm{x}) = \sum_{i=0}^n[logP_{l2t}(z_i|z_{i-1:0},\bm{x}) \\
    + \lambda_r R_r(y_{0:i}, x)],
\end{aligned}
\end{equation}
where $z$ represents the output template of lyric-to-template module, $R_r(y_{0:i}, x)$ is the total rhythm reward of the $i^{th}$ decoding step and $\lambda_r$ represents a hyper-parameter that controls the influence of rhythmic constraint.
Meanwhile, the relationships of tone and structure are added in template-to-melody module:
\begin{equation}
\begin{aligned}
    logP_{t2m}(\bm{y}|\bm{z},\bm{x}) = \sum_{i=0}^n[logP_{t2m}(y_i|y_{i-1:0},\bm{z},\bm{x}) \\
    + \lambda_t R_t(y_{0:i}, x) + \lambda_s R_s(y_{0:i}, x)].
\end{aligned}
\end{equation}
$R_t(y_{0:i}, x)$ and $R_s(y_{0:i}, x)$ are the tone reward and structure reward of the $i^{th}$ step and $\lambda_t$ and $\lambda_s$ refer to hyper-parameters of constraints of tone and structure respectively.

\noindent{\textbf{For SongMASS.}}
SongMASS pre-trains an end-to-end model to generate melody from the given lyrics directly, so the constraints of all three aspects are incorporated to decoding process at once:
\begin{equation}
    \begin{aligned}
    logP(\bm{y}|\bm{x}) = \sum_{i=0}^n[logP_{l2m}(y_i|y_{i-1:0}, \bm{x}) \\
    + \lambda_t R_t(y_{0:i}, x) + \lambda_r R_r(y_{0:i}, x) + \lambda_s R_s(y_{0:i}, x)].
    \end{aligned}
\end{equation}

\section{Experiments}

\begin{table}[t]
    \small
	\centering
	\begin{tabular}{p{1.1cm} | c p{5cm}}
		\toprule
		  & Metric & Description \\
		\midrule
		 & singability & Is the melody smooth and easy to sing? \\
		Melody & diversity & Is the melody novel and not boring? \\
		 & listenability & Is the melody appealing and pleasant? \\
        \midrule
         & intelligibility & Is it easy to understand the lyrics sung by the given melody?\\
		Melody & coherence & Are the melody and lyrics harmonious? \\
		+ Lyrics & structure & Are the melodies and lyrics structures matched? \\
		 & overall & The overall perceptive score for the sample. \\
        \bottomrule
	\end{tabular}
	\caption{Subjective evaluation metrics.}
	\label{tab:subjective_metrics} 
	\vspace{-0.5cm}
\end{table}

\begin{table*}[t]
  \small
	\centering
	\resizebox{\linewidth}{!}{
	\begin{tabular}{clccccccc}
		\toprule
		 \multirow{2}{*}{Language} & \multicolumn{1}{c}{\multirow{2}{*}{Model}} & \multicolumn{2}{c}{Tone} & \multicolumn{2}{c}{Rhythm} & \multicolumn{3}{c}{Structure} \\
		 \cmidrule(lr){3-4} \cmidrule(lr){5-6} \cmidrule(lr){7-9}
		  & & transition~$\uparrow$ & contour~$\uparrow$ & matched s/w positions~$\uparrow$ & matched pauses~$\uparrow$ & PD~$\uparrow$ & DD~$\uparrow$ & MD~$\downarrow$ \\
		\midrule
        \multirow{4}{*}{Chinese} & TeleMelody~\cite{ju2021telemelody} & 0.45 & 0.26 & 0.61 & 0.72 & 0.73 & 0.79 & 2.20\\ 
		& \myname{}~(in TeleMelody) & \textbf{0.61} & \textbf{0.30} & \textbf{0.73} & \textbf{0.79} & \textbf{0.79} & \textbf{0.82} & \textbf{1.53} \\
		\cmidrule(lr){2-9}
        & SongMASS~\cite{sheng2021songmass} & 0.61 & 0.23 & 0.29 & 0.86 & 0.61 & 0.44 & 2.20  \\ 
		& \myname{}~(in SongMASS) & \textbf{0.68} & \textbf{0.36} & \textbf{0.65} & \textbf{0.93} & \textbf{0.69} & \textbf{0.57} & \textbf{0.95} \\ 
		\midrule
		\multirow{4}{*}{English} & TeleMelody~\cite{ju2021telemelody} & - & 0.20 & 0.52 & 0.84 & 0.60 & 0.71 & 1.86\\ 
		& \myname{}~(in TeleMelody) & - & \textbf{0.35} & \textbf{0.69} & \textbf{0.92} & \textbf{0.64} & \textbf{0.73} & \textbf{1.44}\\ 
		\cmidrule(lr){2-9}
		& SongMASS~\cite{sheng2021songmass} & - & 0.25 & 0.44 & 0.82 & 0.61 & 0.54 & 1.54\\ 
		& \myname{}~(in SongMASS) & - & \textbf{0.40} & \textbf{0.54} & \textbf{0.93} & \textbf{0.66} & \textbf{0.58} & \textbf{1.24} \\ 
        \bottomrule
	\end{tabular}
	}
	\caption{Objective results of \myname{} and baseline systems.  Transition score and contour score of tone are calculated as described in Section~\ref{sec:tone_formulation}. For rhythm, we count the average ratio of matched strong/weak positions and pauses respectively. PD, DD and MD are calculated over the repetition parts in melodies.}
	\label{tab:objective_results} 
	\vspace{-0.2cm}
\end{table*}

\begin{table*}[t]
  \small
	\centering
	\resizebox{\linewidth}{!}{
	\begin{tabular}{clcccccccc}
		\toprule
		 \multirow{2}{*}{Language} & \multicolumn{1}{c}{\multirow{2}{*}{Model}} & \multicolumn{3}{c}{Melody} & \multicolumn{4}{c}{Melody + Lyrics} & \multicolumn{1}{c}{\multirow{2}{*}{Average}} \\
		 \cmidrule(lr){3-5} \cmidrule(lr){6-9}
		 & & singability & diversity & listenability & intelligibility & coherence &structure & overall & \\
		\midrule
        \multirow{4}{*}{Chinese} & TeleMelody~\cite{ju2021telemelody} & 2.79 & \textbf{3.29} & 2.92 & 2.94 & 2.51 & 2.73 & 2.78 & 2.85 ($\pm 0.07$) \\
		& \myname{}~(in TeleMelody) & \textbf{3.21} & 3.20 & \textbf{3.16} & \textbf{3.39} & \textbf{2.94} & \textbf{3.16} & \textbf{3.11} & \textbf{3.16 ($\pm 0.06$)} \\
		\cmidrule(lr){2-10}
        & SongMASS~\cite{sheng2021songmass} & 2.74 & \textbf{3.20} & 2.92 & 2.93 & 2.43 & 2.70 & 2.68 & 2.80 ($\pm 0.07$)\\
		& \myname{}~(in SongMASS) & \textbf{3.08} & {3.12} & \textbf{2.98} & \textbf{3.24} & \textbf{2.75} & \textbf{2.89} & \textbf{3.01} & \textbf{3.01 ($\pm 0.06$)} \\ 
		\midrule
		\multirow{4}{*}{English} & TeleMelody~\cite{ju2021telemelody} & 3.27 & 3.45 & 3.19 & 3.78 & 3.32 & 3.21 & 3.23 & 3.35 ($\pm 0.08$) \\
		& \myname{}~(in TeleMelody) & \textbf{3.72} & \textbf{3.47} & \textbf{3.69} & \textbf{4.02} & \textbf{3.61} & \textbf{3.66} & \textbf{3.63} & \textbf{3.68 ($\pm 0.07$)} \\
		\cmidrule(lr){2-10}
		& SongMASS~\cite{sheng2021songmass} & 3.14 & 3.08 & 3.20 & 3.58 & 3.01 & 2.91 & 3.08 & 3.14 ($\pm 0.07$) \\
		& \myname{}~(in SongMASS) & \textbf{3.58} & \textbf{3.26} & \textbf{3.44} & \textbf{3.85} & \textbf{3.20} & \textbf{3.26} & \textbf{3.23} & \textbf{3.40 ($\pm 0.07$)} \\
        \bottomrule
	\end{tabular}
	}
	\caption{The results of subjective evaluation for \myname{} and baseline systems. The average scores are calculated with 95\% confidence intervals.}
	\label{tab:subjective_results} 
	\vspace{-0.2cm}
\end{table*}

\subsection{Experimental Settings}
\subsubsection{Datasets}
\
\newline
We follow the data collection pipeline of TeleMelody~\cite{ju2021telemelody} and SongMASS~\cite{sheng2021songmass} respectively for their training on both English and Chinese song datasets.
Considering that \myname{} requires the pitch contour of a whole sentence, we keep the original punctuations of the lyrics to indicate the intonation of sentences.
Besides, we choose 100 English song samples and 100 Chinese song samples as evaluation datasets to assess the generation quality of \myname{} applied in both TeleMelody and SongMASS. Each sample of the evaluation set does not appear in any of the above training sets. We attach the English and Chinese evaluation sets in \url{https://github.com/microsoft/muzic}.

\subsubsection{System Configuration}
\
\newline
The top-K of stochastic sampling inference is set to 2 and 10 in two modules of TeleMelody and 5 in SongMASS respectively to guide the model to generate melodies with more diversity. 
The temperature is set as 1.2 in both two modules of TeleMelody when studying the influence of hyper-parameters and set as 0.5 when conducting the evaluation.
When apply \myname{} to TeleMelody, we set the hyper-parameters 
${\lambda}_t$, ${\lambda}_r$, ${\lambda}_s$ in constrained decoding as 1.2, 1.5, 1 respectively. And the hyper-parameters  ${\lambda}_t$, ${\lambda}_r$, ${\lambda}_s$ in constrained decoding as 1.5, 1, 1 respectively in SongMASS. The code and configuration can be found in \url{https://github.com/microsoft/muzic}.

\subsubsection{Evaluation Metrics}
\
\newline
We briefly introduce objective and subjective metrics here, more details about the formulation of objective metrics and the instruction of subjective evaluation are listed in Appendix.
\\
\noindent\textbf{Objective Evaluation}
The scores of transition and contour in tone aspect are derived from the rewards in Section~\ref{sec:tone_formulation}. We report the ratios of matched strong/weak positions and matched pause positions for each sample to show the performance of rhythm relationship. As for structure relationship, following~\cite{sheng2021songmass,ju2021telemelody}, we calculate PD (Pitch Distribution Similarity), DD (Duration Distribution Similarity), and MD (Melody Distance) of the repetition parts~\cite{jhamtani2019modeling} in generated melodies. 
\\
\noindent\textbf{Subjective Evaluation}
To evaluate the perceptive quality of the generated melodies, we conduct subjective evaluation and compare \myname{} with the original TeleMelody and SongMASS. We randomly select 15 samples from our test set for each experimental setting and distributed them to 10 participants with music knowledge. All participants are asked to give their scores of each sample from two aspects: 1) the quality of the generated melody itself, and 2) the quality of the overall sample (consider the melody and corresponding lyrics as a whole). The five-point scales (1 for bad and 5 for excellent)~\cite{rec1994p} are used to assess overall generation quality (see Table~\ref{tab:subjective_metrics}). We synthesize the singing voice from the lyrics and melodies using an internal singing voice synthesis tool and provide annotators with the music sheet, lyrics, and singing voice.


\begin{table*}[t]
  \small
	\centering
	\resizebox{\linewidth}{!}{
	\begin{tabular}{lccccccc}
		\toprule
		 \multicolumn{1}{c}{\multirow{2}{*}{Model}} & \multicolumn{2}{c}{Tone} & \multicolumn{2}{c}{Rhythm} & \multicolumn{3}{c}{Structure} \\
		 \cmidrule(lr){2-3} \cmidrule(lr){4-5} \cmidrule(lr){6-8}
		 & transition~$\uparrow$ & contour~$\uparrow$ & matched s/w positions~$\uparrow$ & matched pauses~$\uparrow$ & PD~$\uparrow$ & DD~$\uparrow$ & MD~$\downarrow$ \\
		\midrule
		\myname{} & 0.61 & 0.35 & 0.73 & 0.79 & 0.79 & 0.82 & 1.53 \\
		\ \ --tone relationship &  \cellcolor{gray!25}0.41 & \cellcolor{gray!25}0.25 & 0.74 & 0.80 & 0.93 & 0.89 & 0.98\\ 
		\ \ --rhythm relationship & 0.58 & 0.33 & \cellcolor{gray!25}0.57 & \cellcolor{gray!25}0.69 & 0.76 & 0.93 & 1.41 \\ 
		\ \ --structure relationship & 0.67 & 0.36 & 0.76 & 0.85 & \cellcolor{gray!25}0.74 & \cellcolor{gray!25}0.62 & \cellcolor{gray!25}2.14\\ 
		TeleMelody~\cite{ju2021telemelody} & 0.45 & 0.26 & 0.61 & 0.72 & 0.73 & 0.79 & 2.20 \\
        \bottomrule
	\end{tabular}
	}
	\caption{The objective results of removing tone, rhythm, and structure relationship from \myname{} respectively. The scores of corresponding metrics (gray blocks) decrease.}
	\label{tab:objective_ablation} 
\end{table*}


\subsection{Main Results}
We apply \myname{} in TeleMelody and SongMASS and compare the original version of these two systems with the systems equipped with \myname{}. 
The objective results in Table~\ref{tab:objective_results} show that both TeleMelody and SongMASS are improved after applying \myname{}, verifying the effectiveness of \myname{} and the necessity of incorporating lyric-melody relationships into automatic songwriting.
For original TeleMelody and the one with \myname{}, the gaps of tone scores and structure scores are much larger than those of rhythm, because the musical templates of TeleMelody have already contained the information of rhythmic patterns. In contrast, for SongMASS which does not explicitly capture rhythmic information, the improvement of rhythm scores is significant as tone and structure scores.
Table~\ref{tab:subjective_results} shows the results of subjective evaluation. We can draw the same conclusion as from objective results that \myname{} effectively integrates the relationships between lyrics and melodies into neural network models. By adding rewards of lyric-to-melody relationships to the probability of each token in the decoding stage, \myname{} not only improves the results of objective evaluation but more importantly, enhances the perception of audiences listening to the generated melodies. 

\subsection{Method Analyses}
For simplicity, we performed all method analyses on TeleMelody using the Chinese song dataset. 
\subsubsection{Relationships of Different Aspects}
\
\newline
In this part, we remove tone, rhythm and structure relationship from \myname{} respectively and compare the results. As shown in Table~\ref{tab:objective_ablation}, when removing tone, rhythm or structure relationship, the scores of corresponding objective metrics decrease, which demonstrates the importance of each lyric-melody relationship we introduce. Three relationships have some influence on each other, so \myname{} with all three relationships does not get the highest score for each metric. For example, incorporating the tone relationship influences the structure scores. However, \myname{} achieves a balance between the three aspects, producing the melodies with the highest overall quality.

\subsubsection{Soft Constraints v.s. Hard Constraints}
\
\newline
\myname{} incorporates lyric-melody relationships by adding rewards to the original probabilities of neural network models and considers the rules of the melody itself (learned by neural network models) and lyric-melody relationships at the same time, which is regarded as using ``soft constraints''. If we use ``hard constraints'' in \myname{}, we force the generated melodies to keep in line with the lyrics exactly. As shown in Setting \#2 of Table~\ref{tab:subjective_ablation}, the perceptive quality of the generated melodies is corrupted, indicating that we should balance the rules of the melody itself and lyric-melody relationships, rather than sacrificing the melodic rules to enforce conformity to the lyric-melody relationships.

\begin{table*}[t]
  \small
	\centering
	\resizebox{\linewidth}{!}{
	\begin{tabular}{clcccccccc}
		\toprule
		 \multirow{2}{*}{Setting} & \multicolumn{1}{c}{\multirow{2}{*}{Model}} & \multicolumn{3}{c}{Melody} & \multicolumn{4}{c}{Melody + Lyrics} & \multicolumn{1}{c}{\multirow{2}{*}{Average}} \\
		 \cmidrule(lr){3-5} \cmidrule(lr){6-8}
		 & & singability & diversity & listenability & intelligibility & coherence &structure & overall & \\
		\midrule
        \#1 & TeleMelody~\cite{ju2021telemelody} & 2.79 & \textbf{3.29} & 2.92 & 2.94 & 2.51 & 2.73 & 2.78 & 2.85 ($\pm 0.07$) \\
		\#2 & \ \ + hard constraints & 2.97 & 3.03 & 2.72 & 3.37 & 2.53 & 2.79 & 2.77 & 2.88 ($\pm 0.06$)\\ 
		\#3 & \ \ + conditioned training & 2.96 & 2.92 & 2.79 & 3.08 & 2.67 & 2.72 & 2.75 & 2.84 ($\pm 0.07$) \\
		\#4 & \ \ + re-ranking & 2.89 & 3.08 & 2.74 & 3.32 & 2.59 & 2.79 & 2.81 & 2.90 ($\pm 0.06$) \\
		\#5 & \myname{} & \textbf{3.21} & 3.20 & \textbf{3.16} & \textbf{3.39} & \textbf{2.94} & \textbf{3.16} & \textbf{3.11} & \textbf{3.16 ($\pm 0.06$)}\\
        \bottomrule
	\end{tabular}
	}
	\caption{The results of subjective evaluation for different settings of method analyses. The average scores are calculated with 95\% confidence intervals.}
	\label{tab:subjective_ablation} 
\end{table*}

\begin{figure}[!t]
    \centering
    \includegraphics[width=0.45\textwidth]{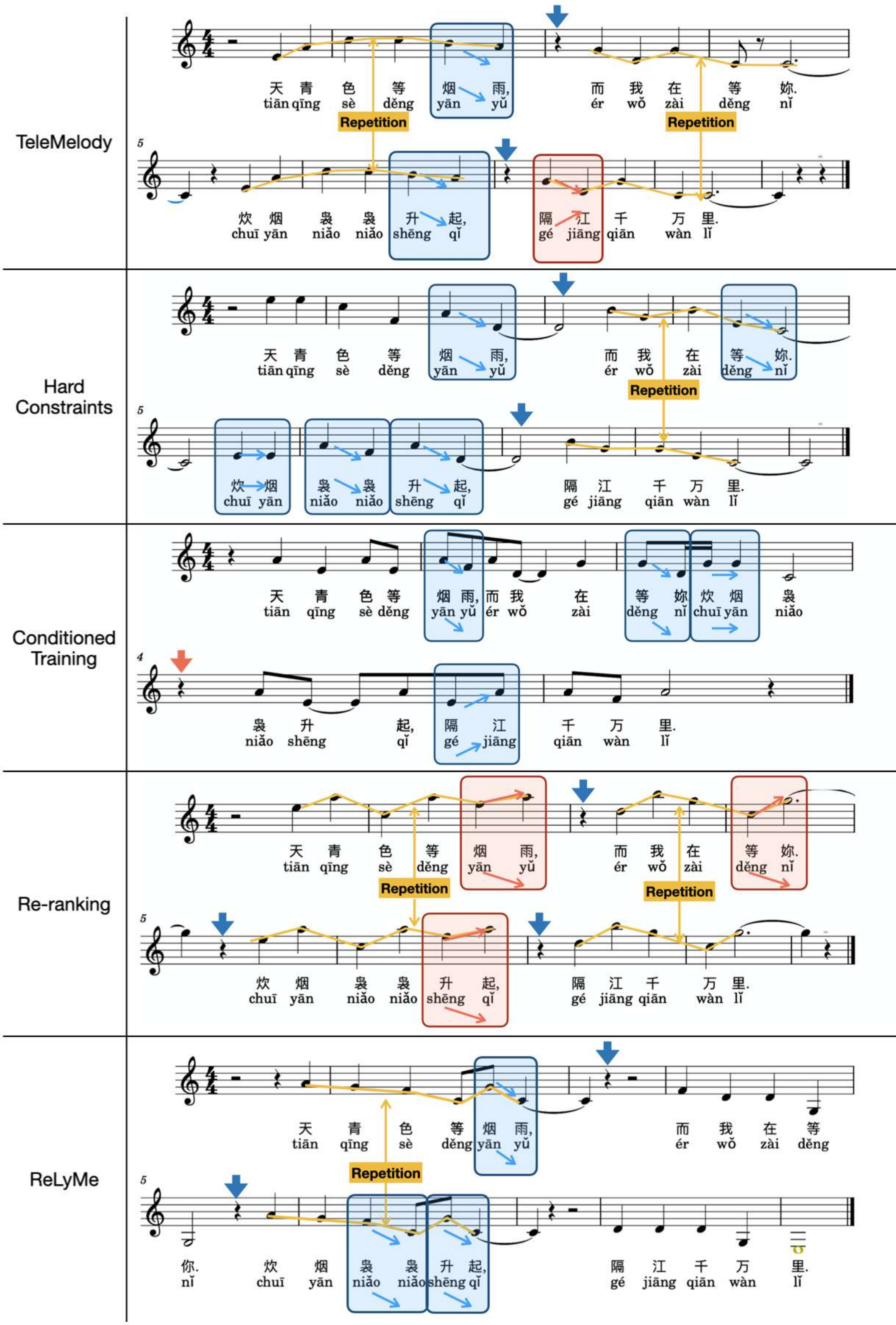}
    \caption{The generated melodies of different settings given the same input lyrics.}
    \label{fig:case_study}
    \vspace{-0.4cm}
\end{figure}


\subsubsection{Conditioned Training \& Re-ranking}
\
\newline
In order to explain why we incorporate the lyric-melody relationships during the decoding process instead of in the training stage or after inference, we compare the results of conditioned training and re-ranking after inference. 
For conditioned training, in the template-to-melody module of TeleMelody, we add the additional pitch information to the input templates. That is, we extract the pitch directions between adjacent notes and add them as a part of the input musical templates to test if they can help with the tone relationships between lyrics and melodies. According to the subjective results in Setting \#3 of Table~\ref{tab:subjective_ablation}, though the intelligibility is higher, other subjective scores decrease, indicating that adding conditions in the training process provides constraints too strong that affect other aspects of the model performance. 
For re-ranking, during the inference, we save several candidates as well as their scores from the model and then re-rank them with the combined scores (the score from the model and the score given by principles of lyric-melody relationships). Then we choose the melody with the highest combined score as the final result. As can be seen in Setting \#4 of Table~\ref{tab:subjective_ablation}, this strategy improves the perceptive quality of generated melodies compared with original TeleMelody, however, performs worse than \myname{}. The results mean that the original lyric-to-melody system can not well capture the lyric-melody relationships, so even the best one of the candidates is not good enough.

\subsubsection{Case Study}
\
\newline
We present the music sheets of samples generated by different methods with the same input lyrics for comparison in Figure~\ref{fig:case_study}.
It is obvious that \myname{} can alleviate most of the tone and rhythm dissonance (red blocks), ensure the harmony between lyrics and melodies (blue blocks), and generate melodies that have the repetition patterns matched with the lyrics (yellow lines). Meanwhile, unlike ``Hard Constraints'', \myname{} retains the flexibility of generated melodies.
More cases and a demo video can be found in \url{https://ai-muzic.github.io/relyme/}.

\section{Conclusion}
In this paper, we propose \myname{}, a method that integrates lyric-melody relationships into neural network models to ensure the harmony between lyrics and melodies. Inspired by musicians, we formulate the principles of lyric-melody relationships from three aspects (tone, rhythm, and structure). \myname{} adjusts the probability of the predicted token at each decoding step according to lyric-melody relationships and can be applied to different lyric-to-melody generation systems without training process. Both objective and subjective results show the effectiveness of \myname{} in improving lyric-to-melody generation by incorporating the relationships between lyrics and melodies. Method analyses further demonstrate the advantage of detailed designs in \myname{}. In the future, we will conduct experiments on more languages and plan to apply the idea of \myname{} to other automatic songwriting tasks, such as melody-to-lyric generation.

\section*{Acknowledgements}

This work was supported by the Key Project of Natural Science Foundation of Zhejiang Province (No.LZ19F020002), the Key R\&D Program of Zhejiang Province (No.2022C03126) and Project of Key Laboratory of Intelligent Processing Technology for Digital Music (Zhejiang Conservatory of Music), Ministry of Culture and Tourism (No.2022DMKLB001).
Thanks are due to Shen Zhou for bringing insights in music.

\bibliographystyle{bib/ACM-Reference-Format}
\bibliography{bib/sample-base}

\newpage

\section*{Appendix}
\label{appendix}

\subsection*{Appendix A. Details of Objective Metrics}
\label{appendix_a}
\textbf{Tone}
\begin{itemize}
    \item Transition: For each adjacent tone pair, we give a score based on the harmony degree of the pitch difference of the corresponding notes. We set the score as 1, 0.5, 0.2, 0 for excellent, good, fair, bad respectively. The transition score is the total score of all adjacent tone pairs divided by the number of all adjacent tone pairs.
    \item Contour: We count the total number of cases in which the pitch contours of melodies match the corresponding lyrics. The contour score is the matched ratio, i.e. the number of matched cases divided by the sentence number. 
\end{itemize}

\noindent\textbf{Rhythm}
\begin{itemize}
    \item Matched s/w positions: We count the total number of matched cases, including the first letter or syllable of the keyword is on a downbeat and the first letter or syllable of the auxiliary word is on a weak beat. Then the number is divided by the total number of auxiliary words and keywords as the matched ratio.
    \item Matched pauses: We count the total number of the cases that do not match, i.e. the REST note or long note is between two syllables within the same word. Then the unmatched number is divided by the total number of inner-word syllables (which means not the first syllable of a word) as an unmatched ratio. The final matched ratio is one minus unmatched ratio.
\end{itemize}

\noindent\textbf{Structure}

Inspired by the evaluation method of SongMASS~\cite{sheng2021songmass}, for the structure objective evaluation, we calculate the pitch and duration distribution similarity and melody distance between the melodies with the same structure.

\subsection*{Appendix B. Instruction of Subjective Evaluation}
\label{appendix_b}
\noindent\textbf{Singability}
\begin{itemize}
    \item How smooth is the melody? (bonus)
    \item Is the melody easy to sing? (bonus)
    \item Are there dissonant pitches or intervals? (penalty)
    \item Are there odd or unexpected rhythms? (penalty)
\end{itemize}
\noindent\textbf{Diversity}
\begin{itemize}
    \item Is the melody creative? Is the melody not easy to predict? (bonus) 
    \item Is the melody monotonous or boring? (penalty)
\end{itemize}
\noindent\textbf{Listenability}
\begin{itemize}
    \item How good is the overall listenability of the melody?
\end{itemize}
\noindent\textbf{Intelligibility}
\begin{itemize}
    \item How easy could you understand the lyrics only depends on the singing clip?
\end{itemize}
\noindent\textbf{Coherence}
\begin{itemize}
    \item Are the melody and the lyric matched in terms of the emotion and the meanings?
    \item Are the prosodies of the melody and the lyric matched? 
\end{itemize}
\noindent\textbf{Structure}
\begin{itemize}
    \item Are there reoccurring themes or motives in the music clip?
    \item What is the degree of harmony between the repetitiveness of the melody and the structure of the lyrics?
\end{itemize}
\noindent\textbf{Overall}
\begin{itemize}
    \item Please give an overall score to this singing clip.
\end{itemize}

\end{document}